# A data-driven quest for room-temperature bulk plastically deformable ceramics

Iwo Słodczyk[1*], Alexander Frisch[2*], Xufei Fang[2], Inna Gitman[1], Fengxian Liu[1*]

[1]Faculty of Engineering Technology, University of Twente, Enschede, the Netherlands

[2]Institute for Applied Materials, Karlsruhe Institute of Technology, Karlsruhe, Germany

**\*Correspondence**

Iwo Słodczyk iwoslodczyk@hotmail.com; Alexander Frisch alexander.frisch@kit.edu; Fengxian Liu f.liu-3@utwente.nl

**Abstract**:

The growing number of ceramics exhibiting bulk plasticity at room temperature has renewed interest in revisiting plastic deformation and dislocation-mediated mechanical and functional properties in these materials. In this work, a data-driven approach is employed to identify the key parameters governing room-temperature bulk plasticity in ceramics. The model integrates an existing dataset of 55 ceramic materials, 38 plastically deformable and 17 brittle, and achieves accurate classification of bulk plasticity.

The analysis reveals several key parameters essential for predicting bulk plasticity: i) Poisson's ratio and Pugh's ratio as macroscopic indicators reflecting the balance between shear and volumetric deformation resistance, and ii) Burgers vector, crystal structure and melting temperature as crystallographic descriptors associated with lattice geometry, slip resistance and thermal stability, and iii) Bader charge as a microscopic measure of bonding character. Together, these parameters define a multiscale descriptor space linking intrinsic materials properties to bulk room-temperature plasticity in ceramics, bridging the gap between empirical ductility criteria and atomistic mechanisms of dislocation-mediated plasticity. While preliminary, this study provides the first systematic, data-driven mapping of the governing factors of ceramic plasticity. The resulting framework establishes a foundation for unifying experimental and computational studies through shared datasets and descriptors, fostering collective progress toward understanding and designing intrinsically ductile ceramics.

# 1. Introduction

The continuous report and discovery of more ceramics capable of *bulk plasticity* mediated by dislocations, mainly at low or room temperature, is creating increasing momentum for a fundamental revisit of the deformation and processing of some ceramics. Here, *bulk plasticity* refers to plastic deformation occurring in macroscopic samples, typically of millimeter size or larger, rather than the localized deformation observed in nano- or microscale mechanical testing. This growing body of evidence stimulates not only the research endeavor for using dislocations to engineer mechanically more resilient and functionally superior ceramics [1], but also contributes to elucidating the dislocation generation mechanisms in novel processing routes such as flash sintering [2], spark plasma sintering [3], and cold sintering [4].

Ceramics are known as inherently strong, chemically inert, and thermally stable, making them ideal candidates for extreme environments such as nuclear reactors, space exploration, and next-generation energy technologies. Their Achilles' heel, however, is brittleness and propensity to crack. In most ceramics, cracks propagate rapidly with little capacity for plastic deformation, resulting in catastrophic failure and significantly limiting their wide range of applications. Achieving intrinsic toughening in otherwise strong ceramics could open new frontiers for high-performance structural materials. On another research forefront, using dislocations to tune versatile physical and chemical properties such as electrical/thermal conductivity, superconductivity, photocatalytic efficiencies, to name a few [5], has shown promising results and potential. However, most of these studies are still limited to a small number of oxides and generating dislocations in some of these model oxides requires nontrivial effort in high-temperature deformation, followed by precise machining and sample preparation [6]. An alternative to this energy and effort-intensive approach is through room-temperature deformation as proposed by the current authors [5, 7].

Encouragingly, recent breakthroughs have challenged the assumption of ceramic brittleness. Recently, 44 ceramics have been summarized which exhibit room-temperature bulk plasticity [8]. These materials span a wide range of lattice structures, slip systems, and chemical bonding environments, showing that ceramic plasticity is indeed possible at low or room temperatures. Yet, the underlying mechanisms that enable such behavior remain poorly understood. Classical theories of plasticity, originally developed for metals and alloys and based on dislocation-mediated deformation, fail to capture the complexity of ceramic plasticity [9]. This is due to ceramic's strong, directional bonding, multicomponent chemistries, and a rich variety of crystal structures.

As emphasized by Thompson and Clegg [10], there exists no universal criterion, whether based on macroscopic elastic constants, such as Pugh's Ratio [11] or Cauchy pressure [12], or on microscopic measures, such as Rice-Thompson's dislocation emission criterion [13, 14], that can reliably distinguish ductile from brittle behaviors across all material classes. While these indicators can sometimes correlate with metallic ductility, they often fail for ceramics, where plasticity mechanisms are far more complex and highly context-dependent. This underlines the need for new frameworks that can capture the multifactorial nature of room-temperature ceramic plasticity, beyond simple scalar descriptors.

At present, the exploration of ceramic plasticity relies heavily on trial-and-error experiments [15-17], which are costly, slow, and fundamentally limited in scope. There is an urgent need for alternative, computational approaches that can **infer patterns from limited data** and provide mechanistic understanding and engineering design guidelines for strong-and-tough as well as functionally superior ceramics.

Data-driven modelling offers a powerful complement to traditional first principles, physics-based approaches, for uncovering hidden correlations between structure and properties based on empirical data. However, two major barriers exist in employing neural network type approaches in the case of room-temperature plasticity in ceramics: 1) data scarcity, up to date, only 44 known plastically deformable ceramics (in bulk tests) have been summarized [8], far too few for conventional machine-learning or deep-learning approaches; 2) undesirability of black-box predictions, many standard machine-learning models lack interpretability to end-users, limiting insight into underlying mechanisms.

To address these two limitations, a fuzzy inference system (FIS) data-driven approach [18, 19] is adopted. Instead of relying on large datasets, FIS models, intrinsically based on fuzzy sets theory [20], exploit the concept of membership functions to handle uncertainty and small datasets. They can capture similarities between subsets of data and generate *"if-then"* rules that describe relationships between intrinsic material parameters and the bulk plastic behaviors. A typical example could read *"if shear modulus is higher than bulk modulus, and the lattice structure is symmetric, then the material is plastically deformable"* (an example rule, simplified here for illustration). Crucially, the derived fuzzy rules are transparent and graphically interpretable, enabling not only prediction but also mechanistic insights.

Fuzzy inference models have a strong track record in prediction applications, including modelling for expert knowledge [20], functioning as universal approximators [18] and, most pertinent, predicting macroscale properties from microscale parameters in metals [19]. These successes give us high confidence that the proposed approach will be able to yield new insights into the mechanisms governing ceramic plasticity.

In what follows, Section 2 details the development of the data-driven model, including construction of the fuzzy inference framework, dataset selection, and parameter sensitivity analysis. Section 3 presents the results and discusses the physical insights and design implications derived from the analysis. Section 4 concludes with key findings and future perspectives.

## 2. Methodology

To investigate the parameters influencing room-temperature bulk plasticity in ceramic materials, a FIS model was implemented in Python and trained on a dataset containing material descriptions and corresponding plasticity metrics for 55 ceramic materials (38 of which demonstrated bulk plasticity at room temperature), referred to as *observations*. The model is an adaptation of an earlier model first introduced by Gitman et al. [19] and further developed by Słodczyk et al. [21, 22] Different combinations of material parameters were investigated, and a sensitivity analysis of the parameters was performed to identify those which have the greatest effect on the prediction accuracy. The sections below first describe how the model is used in conjunction with data as a prediction and analysis method, then the structure of the training and testing data, and finally the methodology of the sensitivity analysis used to evaluate variable importance.

### 2.1 Model development

An illustration of the application of the fuzzy inference model can be seen in Figure 1, where data describing bulk plasticity of ceramic materials (A) is fed into a non-application-specific fuzzy inference model (B) in a process of training and optimization (C). The data consists of ceramic materials descriptors, with one parameter indicating the room-temperature bulk plasticity, referred to as the *consequent*. The remaining parameters report physical, chemical, and engineering properties of a material, which initially are hypothesized to be predictors (individually or collectively) of bulk plasticity, referred to as the *antecedents*. The general fuzzy inference model infers relationships between properties in the data, based on the instances provided in the data, in a process referred to as *training*. Prior to training, the model is agnostic to the data origin and has not been developed with the aim of predicting any specific process. The model is trained (C) and optimized, in a process which determines the prediction error of the model and retrains it using a different set of training parameters, variables which alter the training method referred to as *hyperparameters*, to improve its accuracy. Through attempting a range of hyperparameters, a combination is found which results in an accurate prediction model.

Once the training and optimization step has been completed, the model is then capable of predicting room-temperature bulk plasticity of ceramic materials with a full set of reported physical, chemical, and engineering material characteristics (E). As the next step (F), a sensitivity analysis can be conducted, where the sensitivity of each parameter, and the combinations of parameters, are assessed with regards to the accurate predictions of room-temperature bulk plasticity.

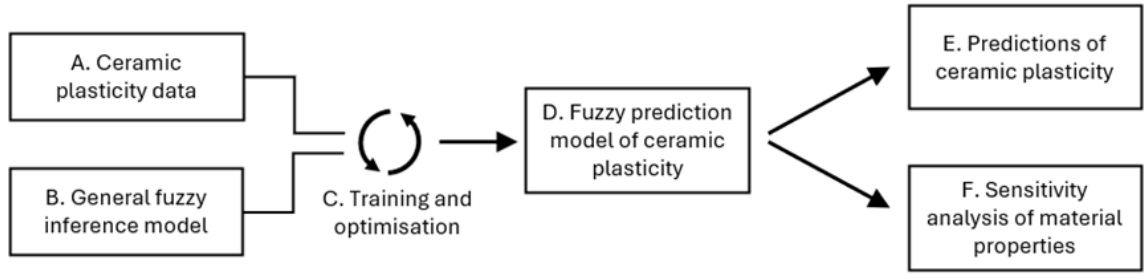

*Figure 1. Flowchart of the fuzzy inference model*

### 2.1.1 Fuzzy concepts and terminology

A short introduction is provided here as a brief reminder of the concepts used in fuzzy methodology. Fuzziness is a measure of the ambiguity of inclusion of an item in a specified group and can be conceptualized as fuzzy sets, which provide a description of agreement (membership) of items with the definition of the set. For example, for a set with the description of "numbers close to 5" the value of 4.9 would have high membership, while the value of 3 would have lower membership, with memberships in the range between 0 (lowest) and 1 (highest). Sets that are not fuzzy are referred to as *crisp* sets and only contain memberships of 0 or 1, crisp values being the common numbers used in most everyday applications. Membership values in a fuzzy set are defined for values of the variable of interest, such as the Poisson's ratio, with the entire range of the variable of interest referred to as the universe of discourse.

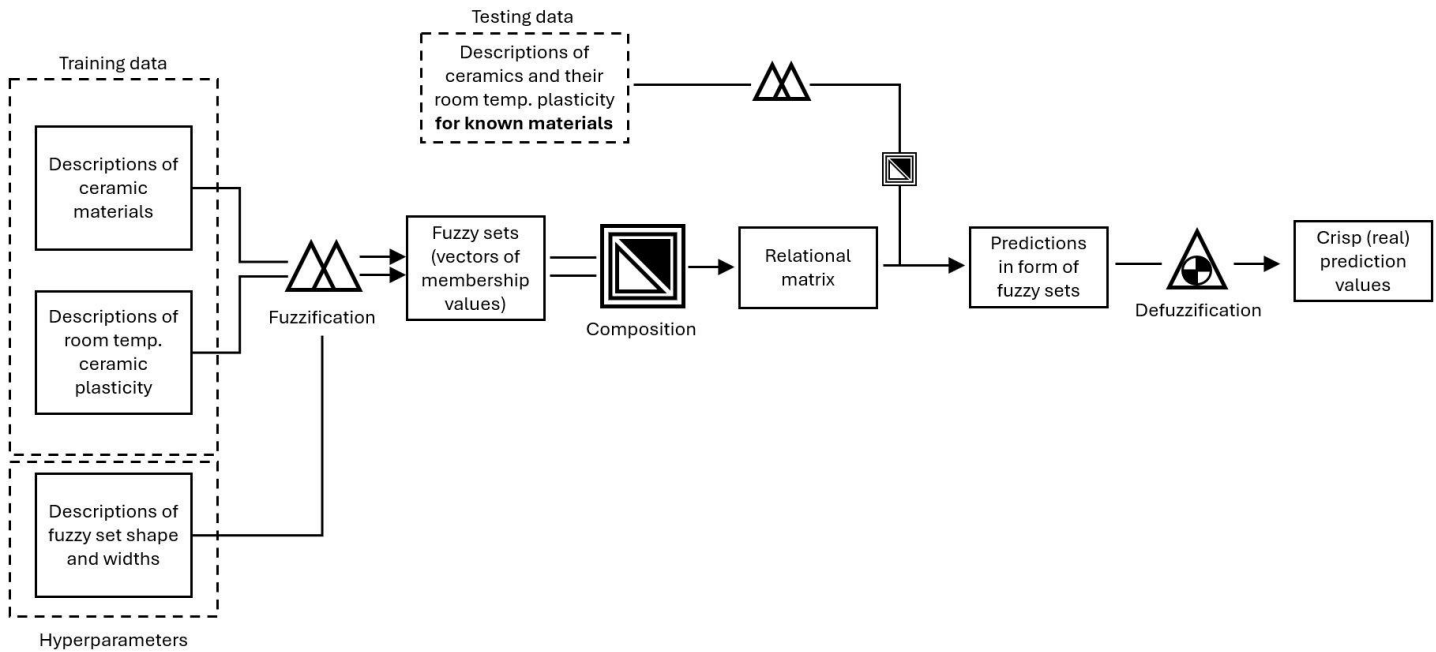

*Figure 2. Data processing steps taken by the fuzzy inference model*

Figure 2 illustrates the data processing steps taken by the fuzzy inference model to achieve a prediction using relationships inferred from the data. The training and prediction process starts with *fuzzification* of the training data using the specification defined in the *hyperparameters*. The *composition* step creates a decision-making unit, the relational matrix, and the testing data is fuzzified and combined with the relational matrix to produce a prediction of the consequent variable in a fuzzy form. Finally, the fuzzy prediction is *defuzzified* to provide a crisp outcome which is the prediction of room-temperature plasticity.

By predicting the plasticity for already known materials and comparing to the known value, a prediction error can be calculated. This error can then be updated by modifying the hyperparameters which dictate the fuzzification process and a hyperparameter value combination resulting in a minimum point of error can be found, thus the entire model can be optimized to produce accurate predictions. Once an accurate model has been achieved, the optimized relational matrix and hyperparameters contain the information necessary to produce future predictions of the same accuracy. The three key processes of fuzzification, composition and defuzzification are further expanded upon in the sections below.

### 2.1.2 Fuzzification

A fuzzy set can be considered as two vectors of equal length, one containing values of the universe of discourse and the other containing corresponding membership values. The process of constructing a fuzzy set by assigning membership values is called fuzzification and is the first fuzzy calculation step seen in Figure 2, taking inputs from the training set and hyperparameters. Many membership functions are available; however, for the purposes of the work presented here, only one is considered – the Gaussian membership function:

$$\mu_i^n = e^{-\frac{1}{2}\left(\frac{v^i - v^n}{w}\right)^2} \tag{1}$$

In the above equation, the membership value $\mu_i^n$ for the value of the universe of discourse $v^i$ for the variable value $v^n$ is calculated using $w$, a width variable decided by the user and contained in the set of hyperparameters. Using Equation 1, a membership value is assigned to each value in the universe of discourse for a given value of a parameter in the training set. Each value in the training set is fuzzified, such that for a dataset containing $m$ materials described using $p$ parameters the number of fuzzy sets calculated is $m \times p$.

### 2.1.3 Composition

The fuzzy inference model calculates predictions using if-then rules, each rule corresponding to an observation in the data, which forms a structure for how the previously mentioned fuzzy sets are applied. The fuzzy rules are expressed in the form:

$$IF\ A_1\ is\ \tilde{A}_1^n\ and\ A_2\ is\ \tilde{A}_2^n\ and\ \dots\ and\ A_i\ is\ \tilde{A}_i^n\ then\ C\ is\ \tilde{C}^n \tag{2}$$

where $A_1$ is variable 1, $\tilde{A}_1^n$ is a fuzzy set for the value of variable 1 for the $n$th material, $C$ is the plasticity descriptor and $\tilde{C}^n$ is the fuzzy set of the value of $C$ for the $n$th material. For the sake of conciseness, the mathematical formulation of the combination of fuzzy sets is omitted here but readers are referred to [22] for a comprehensive examination (including a step-by-step example) of how fuzzy rules are enacted to calculate a prediction.

The combination of fuzzy rules is referred to as composition, and is the second fuzzy calculation step in Figure 2, producing a knowledge base of the relationships between the antecedent variables and the consequent parameter, known as the relational matrix. The relational matrix is the key decision-making unit of the model and its optimization results in more accurate predictions. To calculate a prediction, fuzzy sets of new input values are passed to the relational matrix which combines and transforms them into a fuzzy set of the consequent variable. This set can then be defuzzied to produce a crisp value of the consequent variable, here, a metric of the room-temperature plasticity of the ceramic material.

### 2.1.4 Defuzzification

Transforming the fuzzy set of the consequent into a single, crisp, value is referred to as defuzzification, the last fuzzy calculation step in Figure 2, and can be conducted using a variety of methods. Two common defuzzification methods are centroid and middle of maxima, the former being a weighted average of the whole set while the latter calculates the maximum point of the fuzzy set and its corresponding real-world value. The consequent variable is a categorical variable, sorting materials into either a) plastically deformable or b) not plastically deformable and as such only manifests as one of two discrete values. Therefore, the middle of maxima method has been chosen as more appropriate, whereas the centroid method would be better suited to a continuous variable.

The middle of maxima method finds the universe of discourse value with the highest membership and, should multiple values of equally high membership be found, takes the midpoint between them. Practically, for this application more than one highest membership point was rarely seen, with most calculations resulting in a fuzzy set with a clearly defined peak.

## 2.2 Parameters preparation and data sets setup

The data supplied to the model contains information about the properties of ceramic materials, one of which is a measure of their room-temperature bulk plasticity while the remaining parameters describe physical, chemical, and mechanical properties.

Frisch et al. [8] recently summarized 44 ceramic materials to be plastically deformable in bulk at room temperature. This represents the most up-to-date dataset known to the authors at the time this research was conducted. Among these, only 38 materials contained a complete set of values for the required parameters. To establish a complete prediction range, the 38 plastically deformable observations were supplemented with 17 ceramic materials that are not plastically deformable in bulk at room temperature, resulting in a total of 55 observations. The inclusion of plastically non-deformable ceramics is critical to ensure accurate predictions, without which the model would not be able to differentiate between bulk plastically deformable and non-deformable materials. Table 1 lists all the materials included in the dataset and the parameters used to describe them with their symbols, units and ranges (categories).

### 2.2.1 Candidate parameters potentially indicating room-temperature bulk plasticity

Pugh's ratio $G/K$, where $G$ is the shear modulus and $K$ is the bulk modulus, was included as one of the analyzed parameters due to its continuing use for plasticity predictions [11], although with limited success in non-metallic materials. Pugh's predictions also considered the materials' Burgers vector $b$ and lattice parameter $a$, therefore, these parameters were also included [10, 11]. In ceramics, where dislocations typically dissociate into partials [9], the magnitude of the minimum partial Burgers vector was adopted as the representative value of $b$, as it offers a more physically meaningful measurement for dislocation induced local distortion.

A further proposed predictor for plasticity is the Cauchy pressure, $C_{12} - C_{44}$, which compares the elastic stiffness constants in longitudinal coupling $C_{12}$ and shear stiffness $C_{44}$ to describe the materials' nature of bonding [12]. To further assess the bonding character, the Bader charges, $Q$ of the atomic species were included to measure the number of electrons transferred between ions [23]. Two approaches were used to represent the Bader charge: (i) the Bader charge of the A-site element ($Q_A$) for stoichiometric prototypes of A, AB and ABO$_3$, and (ii) the sum of the positive Bader charges ($Q_{pos}$) over all cations, for example taking the sum of the A and B elements for ABO$_3$, thus giving a more complete characterization of the charge associated with the compound considered as a whole.

Additionally, the Peierls stress is utilized as a measure for lattice resistance against dislocation glide, per definition at 0 K. As there is no universal description for the Peierls stress, that would fit the dislocation behavior of every crystal structure, common constituents of theoretical description of the Peierls stress, specifically Poisson's ratio $\nu$ and slip plane spacing $d$, were chosen here [10].

The melting temperature $T_m$ was introduced as a measure for the materials' thermal stability. The crystal structure ($\zeta$) parameter is a descriptive, categorical parameter which represents the crystal structures of Cesium Chloride, Fluorite, Perovskite, Rock Salt, Sphalerite and Wurtzite, denoted by discrete categories of 1, 2, 3, 4, 5 and 6, respectively. This allows the model to assess whether the knowledge of the crystal structure is necessary to predict plasticity. Note that, certain crystal structures, such as Spinel (AB$_2$O$_4$), were not included due to insufficient or non-consistent data across the selected parameters. The fuzzy inference framework is, however, inherently flexible and can be readily extended to incorporate new or updated datasets as they become available.

Finally, Table 1 presents the consequent, which is also categorical and classifying the compounds as either *bulk plastically deformable at room temperature* (1) or *non-deformable at room temperature* (2). Here, a compound is defined as bulk plastically deformable at room temperature when literature reports demonstrate that dislocation-mediated plastic deformation was shown to precede brittle fracture in bulk-scale uniaxial compression experiments conducted at room temperature. Conversely, plastically non-deformable ceramics were identified from published works, in which the dislocation-mediated bulk plastic deformation specifically required thermal activation, whilst the materials fracture a brittle manner at room temperature.

*Table 1. Dataset of materials used to train and test the fuzzy inference mode, along with the parameters under investigation*

| Materials | | | | | |
|---|---|---|---|---|---|
| Cesium Chloride | Fluorite | Perovskite | Rock Salt | Sphalerite | Wurtzite |
| CsBr[†], CsI[†], TlBr[†] | CaF$_2$[†], BaF$_2$[†], ZrO$_2$[†], UO$_2$, Li$_2$O | SrTiO$_3$[†], KNbO$_3$[†], KTaO$_3$[†], NaNbO$_3$, CaTiO$_3$, BaTiO$_3$ | LiF[†], LiCl[†], LiBr[†], NaF[†], NaCl[†], NaBr[†], NaI[†], KCl[†], KBr[†], KI[†], RbCl[†], RbI[†], MgO[†], CaO[†], CoO[†], NiO[†], SrO[†], AgCl[†], AgBr[†], PbS[†], PbTe[†], TiC, TiN, ZrC, ZrN, NbC, TaC, VC | ZnS[†], ZnSe[†], ZnTe[†], CdTe[†], CuCl[†], CuBr[†], GaAs, InP | ZnO[†], CdS[†], CdSe[†], GaN, w-BN |
| Antecedent (input) variables | | | | | | | | | | | |
| Shear Modulus | Poisson's ratio | Lattice parameter | Melting Temperature | Elastic Constant | Elastic Constant | Bader Charge for A | Crystal Structure | Minimum Burgers vector | Slip plane spacing | Total positive charge | Cauchy pressure | Pugh's ratio |
| $G$ | $\nu$ | $a$ | $T_m$ | $C_{12}$ | $C_{44}$ | $Q_A$ | $\zeta$ | $b_{min}$ | $d$ | $Q_{pos.}$ | $C_{12} - C_{44}$ | $\frac{G}{K}$ |
| GPa | - | Å | K | GPa | GPa | e | - | Å | Å | e | GPa | - |
| 4.6 - 383 | 0.12 - 0.42 | 2.6 - 7.1 | 700 - 4100 | 3 - 185 | 3 - 330 | 0.43 - 2.6 | (1,2,3,4,5,6) | 0.85 - 5.2 | 2.2 - 7.1 | 0.43 - 4.0 | -200 - 100 | 0.15 - 1.0 |
| Consequent (output) variable | | | | | | | | | | | | |
| Room-temperature bulk plasticity category (bulk plastically deformable-1 or bulk plastically nondeformable-2) | | | | | | | | | | | | |

[†]Ceramics marked with [†] are bulk plastically deformable at room temperature that were included in the training and testing datasets. Ceramics without [†] are plastically non-deformation.

2.2.2 Data pre-processing: alternative parameters sets

To minimize interdependence between variables, the parameters discussed in Section 2.2.1 were (re-)grouped to eliminate correlated or mathematically related variables, resulting in four different alternative parameter sets, as shown in Table 2. These four sets are constructed based on exclusionary parameter groups. The first exclusion addressed the interdependence among shear modulus $G$ and Poisson's ratio $\nu$ and Pugh's ratio $G/K$. Since both $G$ and bulk modulus $K$ are related through $\nu$ under standard isotropic elasticity assumptions, Pugh's ratio was not included in any Parameter sets containing both $G$ and $\nu$. The Cauchy pressure, $C_{12}$-$C_{44}$, is directly dependent on its constituent elastic constants; hence, it was not included together with either $C_{12}$ or $C_{44}$ in the same parameter set. Although O. Senkov et. al. [24] reported that Pugh's ratio and Cauchy pressure are equivalent for materials with cubic crystal structures, both Cauchy pressure (or its components $C_{12}$ and $C_{44}$) were retained together with Pugh's ratio (or $G$ and $\nu$) in the Parameter sets to assess their effects across different crystal structures.

Another exclusionary grouping involved the lattice parameter $a$, and the minimum Burgers vector magnitude $b_{min}$, and the interplanar spacing $d$, given their inherent crystallographic dependence. Finally, the parameters $Q_A$ and $Q_{pos}$ were treated as mutually exclusive, as both describe the Bader charge characteristics of the compound, the former representing the charge of the A-site element and the latter the total positive charge of all cations. Parameter sets 1 through 4, thus, represent a progressive refinement of the variable space: Parameter set 1 includes more fundamental parameters, while Parameter set 4 includes more derived or compound parameters.

*Table 2. Selected datasets*

| Parameter set | Parameters |
|---|---|
| 1 | $G, \nu, a, T_m, C_{12}, C_{44}, Q_A, \zeta$ |
| 2 | $G, \nu, b_{min}, T_m, C_{12}, C_{44}, d, Q_{pos.}, \zeta$ |
| 3 | $G, \nu, b_{min}, T_m, (C_{12} - C_{44}), d, Q_{pos.}, \zeta$ |
| 4 | $\frac{G}{K}, b_{min}, T_m, (C_{12} - C_{44}), d, Q_{pos.}, \zeta$ |

### 2.2.3 Training and testing data selections

A careful selection was made of the observations included in the sets of data required to train and optimize the model, the training and the testing data seen in Figure 2. Firstly, observations containing the maximum and minimum of the range for each variable were included in the training dataset. This was done so that no input from the testing set would lie outside of the prediction range of the model, beyond which the model ceases to be accurate. Secondly, observations were added to the training set which were closest to the midpoint, 25th percentile and 75th percentile of the range for each variable. This consideration helps to produce a more even distribution of fuzzy sets, ensuring large gaps are not created between the fuzzy rules for each parameter.

The observations included in the training and testing sets were curated so that each dataset included at least one compound of each crystal structure and at least one deformable and one non-deformable observation for each structure (except for the Cesium Chloride structure which featured only deformable observations). Finally, observations were chosen at random and added to the training set to create a 50:50 training-testing data split from the 55 available observations, which was chosen to give a harsher test than more conventional 80:20 ratios, resulting in 28 observations in the training set and 27 observations in the testing set.

The fuzzy inference model was trained and optimized for each of the four datasets in Table 2, creating four different prediction models. The optimization algorithm, based on Bayesian optimization (Python Bayesian Optimizaion package) uses Gaussian processes to describe the error function of the fuzzy inference model. 1000 iterations were used to minimize the error and by comparing the final accuracy of each model a ranking was established, suggesting which of the four combinations of parameters best predicts room-temperature bulk plasticity of ceramics.

Note here, that while the plasticity metric values of the testing dataset are known, a key use of the model is in predicting room-temperature bulk plasticity for materials with unknown values of this parameter. Used in such a way, the model can predict plasticity for new materials and help guide the design of ceramic materials with desired plasticity.

## 2.3 Sensitivity analysis methodology

Once a predictive model and, hence, a set of predictive parameters enabling the highest accuracy is identified, a sensitivity analysis, allowing for an assessment of the relative importance of parameters in the dataset, can be performed. The outcome of this sensitivity study will help determine which parameters are most critical for accurate predictions of room-temperature bulk plasticity. This study will also help achieve a better understanding of the parameters affecting plasticity, and future experimental testing can then be directed and made more efficient.

To do so, each parameter (variable) was considered individually and for each fuzzy rule (see Section 2.1), the relational matrix of that variable was replaced with one entirely populated by membership values equal to 1. This has the effect of making the model insensitive to changes in the chosen variable and removing any nuance that investigated variable brought to the prediction calculation. This way, by comparing the effect on the predicted values of bulk plasticity before and after replacing the relational matrix, a measure of the importance of the variable in question to the accuracy of the overall calculation can be determined. This process was performed for each variable individually, keeping the relational matrices of the other parameters unaltered and the consequent fuzzy sets were again defuzzified using middle of maxima[1].

---

[1] Note that both the middle of maxima and the centroid defuzzification method were analysed. The centroid method was seen to produce higher values of error in its predictions; however, it was more sensitive to changes in parameters, allowing for a clearer picture of the importance of each variable.

# 3. Results and discussions

## 3.1 Modeling predications and accuracy analysis

Table 3 shows the mean absolute percentage error (MAPE) achieved by the FIS model when trained and optimized for each of the four parameter sets (see Table 2). Each dataset resulted in a reasonably good to excellent accuracy, with parameter sets 2 and 4 having the lowest error, followed by set 1 and finally set 3 having the highest error. The lowest MAPE values of 0% are highly promising, and even the highest error of 3.75% represents strong performance given the small and diverse dataset.

However, these results must be interpreted within the context of the methodology. The bulk plasticity consequent parameter is categorical (plastically deformable or non-deformable) rather than continuous, allowing for a theoretical, perfect (0% MAPE) agreement between predications and reference values. 0% error should not be expected for a continuous consequent variable.

Putting these prediction errors in perspective, within a test of 27 materials, a single misclassification, e.g. identifying one non-deformable material as deformable, results in a MAPE of 1.85%. Again, the same level of accuracy should not be expected when predicting for a more nuanced, continuous plasticity parameter. For prediction instances outside of the range of parameters, a sharp increase in error is expected, and the model should be considered to be an interpolation rather than extrapolation prediction model.

*Table 3. Mean absolute percentage error (MAPE) values for a fuzzy inference model trained on datasets Parameter set 1-4*

| Parameter set | 1 | 2 | 3 | 4 |
|---|---|---|---|---|
| MAPE (%) | 2 | 0 | 3.57 | 0 |

The results presented in Table 3 demonstrate the robustness and internal consistency of the FIS framework, which is supported by the low (and repeated 0%) MAPE values across all four datasets.

## 3.2 Sensitivity analysis and key parameters identification

To further determine which parameter set controls model performance, sensitivity analyses were carried out for the two most accurate Parameter sets (2 and 4). Table 4 lists the corresponding changes in MAPE when each variable was obscured, indicating its influence on prediction accuracy. The positive values of the increase in MAPE are indicative of the variables which were critical for the configuration of the model trained on that parameter set; whereas, zero values of MAPE increase point to those variables which have no effect on the accurate prediction of plasticity. Therefore, only the parameters associated with non-zero increase in MAPE were utilized by the model to predict room temperature bulk plasticity.

For parameter set 2, Poisson's ratio ($\nu$), Bader charge ($Q_{pos}$), and crystal structure ($\zeta$) exhibited the highest sensitivity, while shear modulus ($G$), minimum Burgers vector ($b_{min}$), melting temperature ($T_m$) and elastic constants ($C_{12}$ and $C_{44}$) showed negligible influence. In contrast, parameter set 4 identifies Pugh's ratio ($\frac{G}{K}$), Bader charge ($Q_{pos}$), minimum Burgers vector ($b_{min}$), and melting temperature ($T_m$) as the dominant variables, while crystal structure ($\zeta$) and Cauchy pressure ($C_{12} - C_{44}$) played minor roles. In both cases, the slip plane spacing ($d$) shows negligible influence, suggesting the potential redundancy of this parameter.

It should be noted that not all the critical parameters are the same for parameter set 2 and 4, with disagreement observed for $T_m$ and $\zeta$. These differences indicate that the two models adopt alternative but equally effective descriptor spaces for predicting bulk plasticity. For example, parameter set 2 relies on knowledge of the crystal structure, exhibiting zero sensitivity to other structural parameters such as the Burgers vector; whereas, parameter set 4, which is not sensitive to crystal structure, compensates through dependence on the Burgers vector and melting temperature. This cross-link between related parameters in these two different models provides valuable clues about potential parameter compensation within the same category, see below, which could be further investigated to refine the understanding of plastic deformation mechanisms in ceramics.

*Table 4. Mean absolute percentage error (MAPE) and single variable sensitivity for the PS2 and PS4 datasets*

| Parameter set 2 | | Parameter set 4 | |
|---|---|---|---|
| Parameter investigated | Sensitivity | Parameter investigated | Sensitivity |
| **Mean absolute percentage error (MAPE) (%)** | | | |
| Base | 0 | Base | 0 |
| **Increase in MAPE from Base (%)** | | | |
| $G$ | 0 | $\dfrac{G}{K}$ | 5.36 |
| $\nu$ | 9.26 | | |
| $b_{min}$ | 0 | $b_{min}$ | 5.36 |
| $T_m$ | 0 | $T_m$ | 5.36 |
| $d$ | 0 | $d$ | 0 |
| $C_{12}$ | 0 | $C_{12} - C_{44}$ | 0 |
| $C_{44}$ | 0 | | |
| $Q_{pos.}$ | 3.7 | $Q_{pos.}$ | 10.71 |
| $\zeta$ | 1.85 | $\zeta$ | 0 |

Despite their differences, both parameter sets reveal consistent trends that align with empirical understanding of ceramic plasticity. **Poisson's ratio** and **Pugh's ratio** emerge as key indicators, both capturing the balance between resistance to volumetric and shear deformation, representing parameters long used to differentiate ductile from brittle behavior in solids, particularly in metals [10]. Their appearance among the key parameters confirms that the model predictions are broadly consistent with empirical correlations previous proposed [11].

However, as discussed by Thompson and Clegg [10], such simple macroscopic elastic constant ratios, while convenient because they are derived from easily measurable or calculable quantities, are intrinsically limited when used in isolation. They capture general trends within a given class of materials, where crystal structures and deformation modes remain similar (e.g., pure metals), but fail to account for atomic-scale barriers and crystallographic factors that govern dislocation nucleation and motion. As a result, their transferability across structurally and chemically diverse ceramics is limited.

In this context, the present model moves beyond the constraints of traditional criteria by incorporating parameters that explicitly represent lattice geometry and bonding character. The inclusion of **crystal structure**, **Burgers vector**, and **melting temperature** provides complementary physical insights. The crystal structure defines the possible slip systems, as electrostatic repulsion between ions constrains which dislocation glide planes and directions are energetically favorable. It, therefore, influences the feasible Burgers vectors, whose magnitudes govern the local lattice distortion and directly influence the elastic energy of dislocations (proportional to $b^2$ under isotropic elasticity). The melting temperature, tough less directly linked to dislocation behavior, serves as descriptor for overall bond strength and lattice stability, as well as the thermal activation energy associated with dislocation nucleation and motion. Together, these parameters extend the predictive framework beyond simple elastic moduli, addressing structural factors that traditional ductility criteria omit, and thereby improving the transferability of plasticity predictions across material classes.

Finally, **Bader charge** ($Q_{pos.}$) consistently emerges as a universally important variable across all Parameter sets. The Bader charge quantifies the redistribution of electronic charge among atoms, offering microscopic insights into bonding characters [23, 25]. A small charge transfer between atoms indicates more delocalized, metallic-like bonding, which facilitates dislocation glide and enhances plasticity. Conversely, a large charge transfer corresponds to localized ionic or covalent bonding, which suppresses slip and promotes brittleness. This parameter thus provides a direct link between electronic structure and plastic deformation behavior. Its consistent identification across different Parameter sets highlights the fundamental role of electronic bonding characteristics in enabling room-temperature bulk plasticity in ceramics, an area that demands further theoretical and computational investigation to better understand dislocation mechanisms in ceramics and to guide the design of ductile ceramic materials.

The combined insights from these descriptors establish a multi-dimensional parameter space that integrates macroscopic elasticity, microscopic crystallographic geometry and electronic bonding. Within this space, each parameter family contributes a distinct physical perspective: elastic constants capture the macroscopic mechanical responses; crystallographic parameters describe the lattice geometry, slip resistance, thermal stability; and the electronic parameter reveals the bonding characteristics that play an important role in dislocation nucleation and motion. This multiscale parameter space bridges the gap between empirical ductility criteria and atomistic mechanisms, constructing a transferrable and physically interpretable framework for predicting bulk plasticity in ceramics.

## 4. Conclusions and Perspective

Ceramics exhibiting room-temperature bulk plasticity are no longer scarce but still limited in number. Recent reports demonstrate that such materials span diverse crystal structures and chemical bonding environments. Extending this materials box beyond the currently known cases requires approaches that go beyond experimental trial-and-error explorations. In this context, the data-driven analysis provides a powerful means to accelerate understanding and discovery.

In the present work, a FIS model was employed to identify key material parameters governing room-temperature bulk plasticity in ceramics. Unlike conventional machine-learning approaches that require large dataset and produce black-box prediction, the FIS model offers transparency and robustness in handling limited dataset, making it particularly suited for extract patterns with limited data availability. By integrating a dataset of 55 ceramic materials, the model successfully classified plastically deformable in bulk and non-deformable at room temperature through an interpretable, data-driven framework.

From our analysis, several key parameters are identified as most relevant to dominant bulk plasticity at room temperature: Poisson's ratio and Pugh's ratio, as a macroscopic indicator of the difference between volumetric and shear deformation resistance; Burgers vector, melting temperature and crystal structure reflecting crystallographic factors such as lattice geometry, slip resistance and thermal stability; and the total positive Bader charge as a microscopic indicator of electronic bonding character. Together, these quantities define a descriptor space that links intrinsic material properties to the likelihood of plastic deformation. Narrowing the focus to these parameters offers a rational pathway for systematic experimental and computational investigations to advance understanding of plastic deformation in "ductile" ceramics.

While this study is preliminary and provides a soft categorization of the relative importance of different parameter groups, it represents the first systematic data-driven effort to map the governing factors of ceramic's bulk plasticity. The resulting framework establishes a foundation for unifying experimental and computational studies through shared datasets and common descriptors, fostering collective progress toward revealing the fundamental plastic deformation mechanisms. Building upon this foundation, future efforts should aim to develop quantitative and continuous descriptors of bulk plastic deformability at room temperature, moving beyond the current binary classifications into a continuous scale that captures the degree of plastic responses. Such descriptors will allow a deeper understanding of how elastic, structural, and electronic factors cooperate to enable bulk room-temperature plasticity, ultimately guiding the design of next-generation ceramics that are both strong and intrinsically plastically deformable at room temperature.

Importantly, the proposed framework not only interpret *known* ceramic materials but also has the potential to predict compounds with *unknown* plastic deformability, extending its utility from descriptive analysis to predictive discovery. More broadly, this study establishes a generalizable framework applicable to explore other complex material systems where data are limited and physical interpretability is essential.

**Author contributions:**

**Iwo Słodczyk**: writing—original draft, conceptualization, methodology, investigation, writing—review and editing; **Alexander Frisch**: writing—original draft, conceptualization, investigation, writing—review and editing; **Xufei Fang**: writing—original draft, conceptualization, investigation, writing—review and editing, supervision, funding acquisition; **Inna Gitman**: conceptualization, methodology, investigation, writing—reviewing and editing, supervision; **Fengxian Liu**: writing—original draft, conceptualization, investigation, writing—reviewing and editing, supervision, funding acquisition.

**Declaration of competing interest**

The authors declare that they have no known competing financial interests or personal relationships that could have appeared to influence the work reported in this paper.

**Acknowledgments**

F. Liu would like to acknowledge the financial support from the University of Twente (ET Crazy Research 2023-BettEr). A. Frisch and X. Fang thank the financial support from European Research Council (ERC Starting Grant, project MECERDIS, grant number 101076167).